\documentclass[twoside,english]{IEEEtran}
\usepackage[T1]{fontenc}
\usepackage[latin9]{inputenc}
\usepackage{amsmath}
\usepackage{graphicx}
\usepackage{amssymb}

\usepackage{color, graphicx}
\usepackage{cite}
\usepackage{epsfig, psfrag}
\usepackage{subfigure}
\newtheorem{theorem}{Theorem}[section]

\def\Yi{\mathbf{Y}_{\hspace{-1pt}i}(t)}
\def\SYi{\mathcal{S}_{\mathbf{Y}_{\hspace{-1pt}i}}(f)}

\usepackage{babel}

\begin{document}

\title{Communication in a Poisson Field of Interferers -- Part~II: Channel
Capacity and Interference Spectrum}

\author{Pedro C.\ Pinto, \IEEEmembership{Student Member, IEEE,} and Moe
Z.\ Win, \IEEEmembership{Fellow, IEEE}%
\thanks{Manuscript received November~16, 2007; accepted Mar~18, 2008. The
editor coordinating the review of this paper and approving it for
publication is D.~Dardari. This research was supported by the Portuguese
Science and Technology Foundation under grant SFRH-BD-17388-2004,
the Charles Stark Draper Laboratory Robust Distributed Sensor Networks
Program, the Office of Naval Research Young Investigator Award N00014-03-1-0489,
and the National Science Foundation under Grant ANI-0335256. This
paper was presented, in part, at the IEEE Global Telecommunications
Conference, San Francisco, CA, November 2006.%
}%
\thanks{P.~C.~Pinto and M.~Z.~Win are with the Laboratory for Information
and Decision Systems (LIDS), Massachusetts Institute of Technology,
Room~\mbox{32-D674}, 77~Massachusetts Avenue, Cambridge, MA 02139,
USA (\mbox{e-mail}: \texttt{ppinto@mit.edu}, \texttt{moewin@mit.edu}). %
}%
\thanks{Digital Object Identifier 10.1109/TWC.2008.XXXXXXX%
}}

\maketitle

\thispagestyle{empty}

\newpage

\setcounter{page}{1}
\begin{abstract}
In Part~I of this paper, we presented a mathematical model for communication
subject to both network interference and noise, where the interferers
are scattered according to a spatial Poisson process, and are operating
asynchronously in a wireless environment subject to path loss, shadowing,
and multipath fading. We determined the distribution of the aggregate
interference and the error performance of the link. In this second
part, we characterize the capacity of the link subject to both network
interference and noise. Then, we put forth the concept of \emph{spectral
outage probability} (SOP), a new characterization of the aggregate
radio-frequency emission generated by communicating nodes in a wireless
network. We present some applications of the SOP, namely the establishment
of spectral regulations and the design of covert military networks.
The proposed framework captures all the essential physical parameters
that affect the aggregate network emission, yet is simple enough to
provide insights that may be of value in the design and deployment
of wireless networks.\end{abstract}
\begin{keywords}
Spatial distribution, Poisson field, aggregate network emission, channel
capacity, spectral outage, stable laws.
\end{keywords}

\section{Introduction\label{sec:Introduction}}

\PARstart{T}{he}  application of the spatial Poisson process to cellular
networks was investigated in \cite{Sou:92}, and later advanced in
\cite{IloHatVen:98,YanPet:03,GovAntBliSta:05}. However, these studies
focus mostly on error performance metrics, and do not attempt a characterization
of the channel capacity and interference spectrum. Furthermore, they
often ignore random propagation effects (e.g.~shadowing and fading)~\cite{Sou:92};
assume perfect synchronization between different interferers at the
symbol or slot level~\cite{YanPet:03}; or restrict the node locations
to a disk in the two-dimensional plane~\cite{GovAntBliSta:05,SalZan:07},
which complicates the analysis and does not provide useful insights
into the effects of network interference. In \cite{GioChiWin:J05,GioDar:05},
the authors analyze coexistence issues in narrowband and ultrawideband
networks, but consider only a small, fixed number of interferers.

In Part~I of this paper~\cite{PinWin:J09}, we introduced a framework
where the interferers are scattered according to a spatial Poisson
process, and are operating asynchronously in a wireless environment
subject to path loss, shadowing, and multipath fading~\cite{PinWin:C06,PinWin:J07,PinChoGioChiWin:C06,WinPinGioChiShe:C06}.
Under this scenario, we determined the statistical distribution of
the aggregate interference, and the corresponding error performance
of the link. In this second part, we characterize the capacity of
the link subject to both network interference and noise. Then, we
put forth the concept of spectral outage probability (SOP), a new
characterization of the aggregate radio-frequency (RF) emission generated
by communicationg nodes in a wireless network. Lastly, we quantify
these metrics as a function of important system parameters, such as
the signal-to-noise ratio (SNR), interference-to-noise ratio (INR),
path loss exponent of the channel, and spatial density of the interferers.
Our analysis easily accounts for all the essential physical parameters
that affect the aggregate network emission. Furthermore, the concept
of SOP can be used (e.g.~in commercial or military applications)
to evaluate and limit the impact of network interference on any given
receiver operating in the same frequency band.

This paper is organized as follows. Section~\ref{sec:System-model}
briefly reviews the system model introduced in Part~I. Section~\ref{sec:Channel-Capacity}
analyzes the channel capacity of the system, and presents numerical
examples to illustrate its dependence on important network parameters.
Section~\ref{sec:Spectral-characterization} derives the PSD of the
aggregate interference, introduces the concept of spectral outage
probability, and provides numerical examples of both metrics. Section~\ref{sec:Summary}
summarizes important findings.

\section{Model Summary\label{sec:System-model}}

We briefly review the model introduced in Part~I. As shown in \cite[Fig. 1]{PinWin:J09},
we consider the interfering nodes to be spatially scattered in the
two-dimensional infinite plane, according to a homogeneous Poisson
process with density~$\lambda$ (in nodes per unit area). The random
distance of interfering node~$i$ to the origin is denoted by $R_{i}$.
For analytical purposes, we introduce a \emph{probe link} which is
composed of two nodes: the \emph{probe receiver} (located at the origin),
and the \emph{probe transmitter} (node~$\mbox{$i=0$}$).

In terms of transmission characteristics, we consider that all interfering
nodes employ the same two-dimensional modulation and transmit at the
same power~$P$. For generality, however, we allow the probe transmitter
to employ an arbitrary two-dimensional modulation and arbitrary power~$P_{0}$,
not necessarily equal to that used by the interfering nodes. We consider
that all nodes employ the same symbol rate~$1/T$, but the signal
received from node~$i$ is shifted by a random delay~$D_{i}$, where
$\mbox{$D_{i}\sim\mathcal{U}(0,T)$}$.%
\footnote{We use $\mathcal{U}(a,b)$ to denote a real uniform distribution in
the interval~$[a,b]$.%
} The probe receiver performs coherent demodulation of the desired
signal using a conventional in-phase/quadrature (IQ) detector.

The wireless propagation channel introduces path loss, log-normal
shadowing, and multipath fading. Specifically, the overall effect
of the channel on node~$i$ is accounted for by the random phase~$\mbox{$\phi_{i}\sim\mathcal{U}(0,2\pi)$}$,
and the amplitude factor~$\frac{k\alpha_{i}e^{\sigma G_{i}}}{R_{i}^{b}}$.
The term~$\frac{k}{R_{i}^{b}}$ accounts for the path loss; $\alpha_{i}$
is due to the multipath fading, and has an arbitrary distribution
with $\mbox{$\mathbb{E}\{\alpha_{i}^{2}\}=1$}$; and $e^{\sigma G_{i}}$
is due to the log-normal shadowing, with $\mbox{$G_{i}\sim\mathcal{N}(0,1)$}$.%
\footnote{We use $\mathbb{E}\{\cdot\}$ and $\mathbb{V}\{\cdot\}$ to denote
the expectation and variance operators, respectively. In addition,
we use $\mathcal{N}(\mu,\sigma^{2})$ to denote a real Gaussian distribution
with mean~$\mu$ and variance~$\sigma^{2}$.%
}

In the rest of the paper, we consider the scenario where the location~$\{R_{i}\}_{i=1}^{\infty}$
and shadowing~$\{G_{i}\}_{i=1}^{\infty}$ of the interferers (succinctly
denoted by $\mathcal{P}$), as well as the shadowing~$G_{0}$ affecting
the probe transmitter, remain approximately constant during the interval
of interest. This models a quasi-static scenario where the movement
of the nodes during the interval of interest is negligible. In such
case, we condition the analysis on $\mathcal{P}$ in order to derive
a \emph{capacity outage probability} and a \emph{spectral outage probability},
which are more meaningful than the corresponding $\mathcal{P}$-averaged
metrics.%
\footnote{We implicitly assume conditioning on $\mathcal{P}$ in the rest of
the paper, unless otherwise indicated.%
} Other fast-varying propagation effects, such as multipath fading
due to local scattering, are averaged out in the analysis.

\section{Channel Capacity\label{sec:Channel-Capacity}}

In Part~I of this paper, we focused on error performance metrics.
We now build on the results of Part~I and analyze the capacity of
the link between the probe transmitter and probe receiver in \cite[Fig. 1]{PinWin:J09},
subject to aggregate network interference and additive white Gaussian
noise (AWGN). Unlike the simple AWGN channel, here the capacity depends
on the information available about the channel at the probe transmitter
and receiver. As in Part~I, we assume that the probe receiver can
perfectly estimate the  fading ($\alpha_{0}$ and $\phi_{0}$) affecting
its own link, thus ensuring that coherent demodulation of the desired
signal is possible. The probe transmitter, on the other hand, is not
able to estimate the channel. This corresponds to the scenario where
the receiver has perfect knowledge of the channel side information
(CSI).

\subsection{Capacity Outage Probability\label{sub:Capacity-Outage-Prob}}

We start with the complex baseband characterization of the probe link,
obtained in Part~I by projecting all signals onto a cosine-sine orthonormal
set. Thus, the complex channel output~$\mathbf{Z}$ can be written
as\begin{equation}
\mathbf{Z}=\frac{\alpha_{0}e^{\sigma G_{0}}}{r_{0}^{b}}\mathbf{S}+\mathbf{\widetilde{W}},\label{eq:z-capacity}\end{equation}
where $\mathbf{S}$ is the complex channel input, and $\mathbf{\widetilde{W}}$
is the combined aggregate interference and thermal noise, given by\begin{equation}
\mathbf{\widetilde{W}}=\sum_{i=1}^{\infty}\frac{e^{\sigma G_{i}}\mathbf{X}_{i}}{R_{i}^{b}}+\mathbf{W},\label{eq:W-tilde}\end{equation}
with $\mathbf{W}\sim\mathcal{N}_{\textrm{c}}(0,N_{0}).$%
\footnote{We use $\mathcal{N}_{\textrm{c}}(0,\sigma^{2})$ to denote a circularly
symmetric (CS) complex Gaussian distribution, where the real and imaginary
parts are i.i.d.\ $\mathcal{N}(0,\sigma^{2}/2)$.%
} These are essentially the same baseband equations as those given
in Part~I, except that the transmitted constellation symbol~$a_{0}e^{j\theta_{0}}$
has been replaced by a generic input symbol~$\mathbf{S}$, with an
arbitrary distribution~$f_{\mathbf{S}}(s)$. This emphasizes the
fact that to analyze the channel capacity, we need to maximize the
mutual information over all possible input distributions~$f_{\mathbf{S}}(s)$,
and thus cannot restrict $\mathbf{S}$ to belong to a specific constellation,
such as $\mbox{$M$-PSK}$ or $\mbox{$M$-QAM}$. In addition, we impose
an average energy constraint on the input symbol by requiring that
$\mbox{$\mathbb{E}\{|\mathbf{S}|^{2}\}\leq E_{\mathrm{S}}$}$.

Considering that the interfering nodes are coded and operating close
to capacity, then the signal transmitted by each interferer is Gaussian,
such that $\mathbf{X}_{i}\sim\mathcal{N}_{\textrm{c}}(0,2V_{X})$~\cite{Fos:07}.%
\footnote{Alternatively, we can follow the same approach as in Part~I and argue
that $\mathbf{X}_{i}\sim\mathcal{N}_{\textrm{c}}(0,2V_{X})$ in a
scenario where the interferers employ an arbitrary two-dimensional
modulation (this is the Gaussian approximation introduced in \cite[Eq. (10)]{PinWin:J09}.
In such case, $V_{X}$ is a function of the constellation of the interferers,
as shown in \cite[Eq. (13)]{PinWin:J09}.%
} The resulting aggregate network interference is thus Gaussian when
conditioned on $\mathcal{P}$, and the distribution of $\mathbf{\widetilde{W}}$
in (\ref{eq:W-tilde}) is given by%
\footnote{We use $X\,\mathop\sim\limits ^{{\scriptscriptstyle |Y}}$ to denote
the distribution of $X$ conditional on $Y$.%
}\begin{equation}
\mathbf{\widetilde{W}}\,\mathop\sim\limits ^{{\scriptscriptstyle |\mathcal{P}}}\,\mathcal{N}_{\mathrm{c}}(0,2AV_{X}+N_{0}),\label{eq:W-tilde-normal}\end{equation}
where\begin{equation}
A=\sum_{i=1}^{\infty}\frac{e^{2\sigma G_{i}}}{R_{i}^{2b}}.\label{eq:A-sum}\end{equation}
Note that since $A$ in (\ref{eq:A-sum}) depends on $\mathcal{P}$
(i.e.,~$\{R_{i}\}_{i=1}^{\infty}$ and $\{G_{i}\}_{i=1}^{\infty}$),
it can be seen as a random variable (r.v.) whose value is different
for each realization of $\mathcal{P}$. It was shown in Part~I that
the r.v.~$A$ has a \emph{skewed stable distribution}~\cite{SamTaq:94}
given by%
\footnote{We use $\mathcal{S}(\alpha,\beta,\gamma)$ to denote a real stable
distribution with characteristic exponent~$\mbox{$\alpha\in(0,2]$}$,
skewness~$\mbox{$\beta\in[-1,1]$}$, and dispersion~$\mbox{$\gamma\in[0,\infty)$}$.
The corresponding characteristic function is\[
\phi(w)=\begin{cases}
\exp\left[-\gamma|w|^{\alpha}\left(1-j\beta\mathrm{\, sign}(w)\tan\frac{\pi\alpha}{2}\right)\right], & \alpha\neq1,\\
\exp\left[-\gamma|w|\left(1+j\frac{2}{\pi}\beta\mathrm{\, sign}(w)\ln|w|\right)\right], & \alpha=1.\end{cases}\]
}\begin{equation}
A\sim\mathcal{S}\left(\alpha_{A}=\frac{1}{b},\:\beta_{A}=1,\:\gamma_{A}=\lambda\pi C_{1/b}^{-1}e^{2\sigma^{2}/b^{2}}\right),\label{eq:stableA}\end{equation}
where $\mbox{$b>1$}$, and $C_{x}$ is defined as\begin{equation}
C_{x}\triangleq\begin{cases}
\frac{1-x}{\Gamma(2-x)\cos(\pi x/2)}, & x\neq1,\\
\frac{2}{\pi}, & x=1,\end{cases}\label{eq:Cx}\end{equation}
with $\Gamma(x)=\int_{0}^{\infty}t^{x-1}e^{-t}dt$ denoting the gamma
function. 

\begin{figure}
\begin{centering}
\scalebox{1.1}{\psfrag{S}{$\mathbf{S}$}
\psfrag{W}{$\mathbf{\widetilde{W}}$}
\psfrag{Z}{$\mathbf{Z}$}
\psfrag{e/R}{\large{$\frac{\alpha_{0}e^{\sigma G_{0}}}{r_{0}^{b}}$}}
\includegraphics{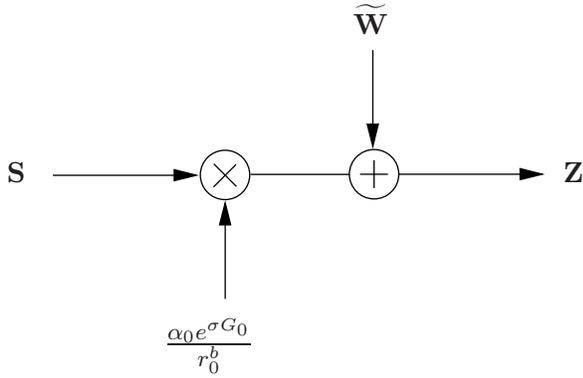}}
\par\end{centering}

\caption{\label{cap:capacitymdl}Channel model for capacity analysis.}

\end{figure}

Because of the conditioning on $G_{0}$ and $\mathcal{P}$, equations
$\mbox{(\ref{eq:z-capacity})-(\ref{eq:A-sum})}$ describe a simple
Gaussian channel depicted schematically in Fig.~\ref{cap:capacitymdl}.
The capacity of this energy-constrained, fast fading channel with
receiver CSI can be written as~\cite{TseVis:05}\[
C=\max_{f_{\mathbf{S}}:\,\mathbb{E}|\mathbf{S}|^{2}\leq E_{\mathrm{S}}}I(\mathbf{S};\mathbf{Z}|\alpha_{0}),\]
where $I(\mathbf{S};\mathbf{Z}|\alpha_{0})$ is the conditional mutual
information between $\mathbf{S}$ and $\mathbf{Z}$ given $\alpha_{0}$.
The optimal input distribution that maximizes the mutual information
is therefore $\mathcal{N}_{\mathrm{c}}(0,E_{\mathrm{S}})$. With this
input distribution,\[
I(\mathbf{S};\mathbf{Z}|\alpha_{0}=\widetilde{\alpha}_{0})=\log_{2}\left(1+\frac{\widetilde{\alpha}_{0}^{2}e^{2\sigma G_{0}}E_{\mathrm{S}}}{r_{0}^{2b}(2AV_{X}+N_{0})}\right)\]
 in bits per complex symbol, and thus we obtain the capacity of the
channel as\begin{equation}
C(G_{0},\mathcal{P})=\mathbb{E}_{\alpha_{0}}\left\{ \left.\log_{2}\left(1+\frac{\alpha_{0}^{2}e^{2\sigma G_{0}}E_{\mathrm{S}}}{r_{0}^{2b}(2AV_{X}+N_{0})}\right)\right|G_{0},A\right\} \label{eq:C-G0-P}\end{equation}
in bits per complex symbol, where we have explicitly indicated the
conditioning of $C$ on the random interferer positions and shadowing.
For a Rayleigh fading channel, $\alpha_{0}^{2}$ is exponentially
distributed with mean~$1$ and we can further express (\ref{eq:C-G0-P})
in terms of the exponential integral function~$\mbox{$\textrm{Ei}(x)=-\int_{-x}^{\infty}\frac{e^{-t}}{t}dt$}$
as\begin{equation}
C(G_{0},\mathcal{P})=-\frac{\exp\left(\frac{\sqrt{2}}{\eta}\right)}{\ln(2)}\textrm{Ei}\left(-\frac{\sqrt{2}}{\eta}\right)\label{eq:C-G0-P-Ei}\end{equation}
in bits per complex symbol, where\begin{equation}
\eta=\frac{e^{2\sigma G_{0}}E_{\mathrm{S}}}{r_{0}^{2b}(2AV_{X}+N_{0})}\label{eq:eta}\end{equation}
is the received signal-to-interference-plus-noise ratio (SINR), averaged
over the fast fading.

In the proposed quasi-static model, the maximum rate of reliable communication
for a given realization of $G_{0}$ and $\mathcal{P}$ is given by
$\mbox{(\ref{eq:C-G0-P-Ei})-(\ref{eq:eta})}$. Such quantity is a
function of the random user positions and shadowing, and is therefore
random. Then, with some probability, $G_{0}$ and $\mathcal{P}$ are
such that the capacity is below the transmission rate~$R$, thus
making the channel unusable for communication at that rate with arbitrarily
low error probability. The system is said to be \emph{in outage},
and the capacity outage probability is\begin{equation}
P_{\textrm{out}}^{\textrm{c}}=\mathbb{P}_{G_{0},\mathcal{P}}\{C(G_{0},\mathcal{P})<R\},\label{eq:Cout-definition}\end{equation}
or, substituting $(\ref{eq:C-G0-P-Ei})$ into (\ref{eq:Cout-definition}),\begin{equation}
P_{\textrm{out}}^{\textrm{c}}=\mathbb{P}_{\eta}\left\{ -\frac{\exp\left(\frac{\sqrt{2}}{\eta}\right)}{\ln(2)}\textrm{Ei}\left(-\frac{\sqrt{2}}{\eta}\right)<R\right\} .\label{eq:Cout-Ei}\end{equation}

\subsection{Numerical Results}

\begin{figure}
\begin{centering}
\scalebox{0.45}{\psfrag{Pcout}{\LARGE{$P_{\textrm{out}}^{\textrm{c}}$}}
\psfrag{SNR (dB)}{\Large{$\mathsf{SNR\;(dB)}$}}
\psfrag{INR=-inf,10,20,30 dB}{\LARGE{$\mathsf{INR=-\infty,10,20,30\;dB}$}}\includegraphics{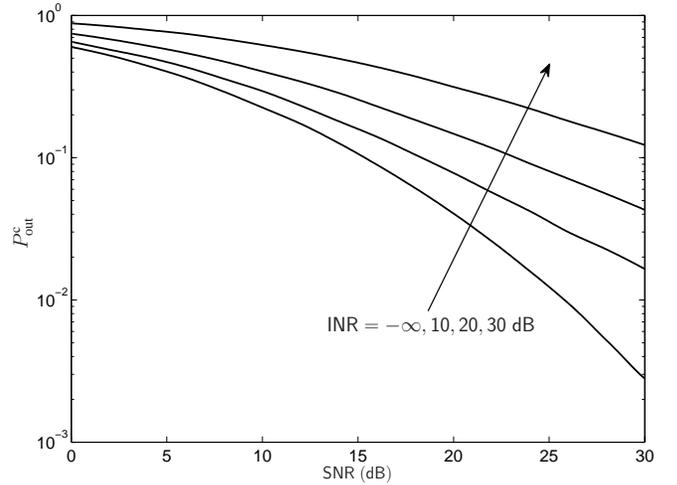}}
\par\end{centering}

\caption{\label{cap:Cout-approx-inr-plot}Capacity outage probability~$P_{\textrm{out}}^{\textrm{c}}$
versus the SNR of the probe link, for various interferer-to-noise
ratios~INR ($\mbox{$R=1\,\textrm{bit/complex symbol}$}$, $\mbox{$\lambda=0.01\,\textrm{m}^{-2}$}$,
$\mbox{$b=2$}$, $\mbox{$r_{0}=1\,\textrm{m}$}$, $\sigma_{\textrm{dB}}=10$).}

\end{figure}

\begin{figure}
\begin{centering}
\scalebox{0.45}{\psfrag{Pcout}{\LARGE{$P_{\textrm{out}}^{\textrm{c}}$}}
\psfrag{rate (bits/complex symbol)}{\Large{\sf{rate $R$ (bits/complex symbol)}}}
\psfrag{la=10e-4,10e-3,10e-2,10e-1}{\LARGE{$\mathsf{\lambda=10^{-4},10^{-3},10^{-2},10^{-1}\, m^{-2}}$}}\includegraphics{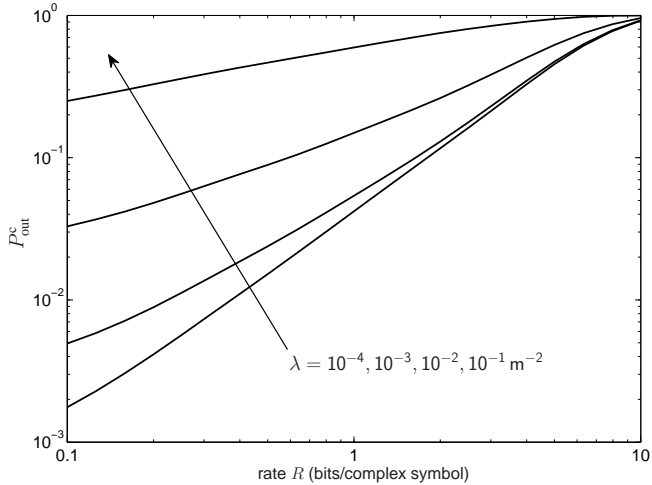}}
\par\end{centering}

\caption{\label{cap:Cout-approx-lambda-plot}Capacity outage probability~$P_{\textrm{out}}^{\textrm{c}}$
versus the transmission rate~$R$, for various interferer spatial
densities~$\lambda$ ($\mbox{$\mathsf{SNR}=\mathsf{INR}=20\,\textrm{dB}$}$,
$\mbox{$b=2$}$, $\mbox{$r_{0}=1\,\textrm{m}$}$, $\sigma_{\textrm{dB}}=10$).}

\end{figure}

Figures~\ref{cap:Cout-approx-inr-plot} and \ref{cap:Cout-approx-lambda-plot}
quantify the capacity outage probability and illustrate its dependence
on the various parameters, such as the signal-to-noise ratio~$\mbox{$\mathsf{SNR}=E_{\mathrm{S}}/N_{0}$}$,
the interference-to-noise ratio~$\mbox{$\mathsf{INR}=E/N_{0}$}$,
and spatial density~$\lambda$ of the interferers. For simplicity,
we consider a case study where all interfering nodes transmit equiprobable
symbols, belonging to a constellation that is symmetric with respect
to the origin of the IQ-plane (e.g.,~$\mbox{$M$-PSK}$ and $\mbox{$M$-QAM}$).
In this particular case, it is shown in \cite[Eq. (13)]{PinWin:J09}
that $\mbox{$V_{X}=E/3$}$, and thus (\ref{eq:eta}) reduces to\begin{equation}
\eta=\frac{e^{2\sigma G_{0}}\mathsf{SNR}}{r_{0}^{2b}\left(\frac{2A}{3}\mathsf{INR}+1\right)}.\label{eq:eta-simplif}\end{equation}
To evaluate the corresponding $P_{\textrm{out}}^{\textrm{c}}$, we
resort to a hybrid approach where we employ the analytical result
given in $\mbox{(\ref{eq:Cout-Ei})-(\ref{eq:eta-simplif})}$, but
perform a Monte Carlo simulation of the stable r.v.~$A$ according
to \cite{ChaMalStu:76}. Nevertheless, we emphasize that the expressions
derived in this paper completely eliminate the need for simulation
of the interferers' positions and waveforms in the network, in order
to obtain the capacity.

\section{Spectral Characterization of the Aggregate Network Emission\label{sec:Spectral-characterization}}

The spectral occupancy and composition of the aggregate radio-frequency
(RF) emission generated by a network is an important consideration
in the design of wireless systems. In particular, it is often beneficial
to know the spectral properties of the aggregate RF emission generated
by all the spatially scattered nodes in the network. This is useful
in commercial applications, for example, where communication designers
must ensure that the RF emission of the network does not cause interference
to other systems operating in overlapping frequency bands. To prevent
interference, many commercial networks operate under restrictions
which often take the form of spectral masks, imposed by a regulatory
agency such as the US Federal Communications Commission (FCC). In
military applications, on the other hand, the goal is ensure that
the presence of the deployed network is not detected by the enemy.
If, for example, a sensor network is to be deployed in enemy territory,
then the characterization of the aggregate network emission is essential
for the design of a covert system.

In Part~I and in Section~\ref{sec:Channel-Capacity} of this paper,
we derived the error probability and capacity of a link subject to
both network interference and thermal noise. We now determine the
PSD of the aggregate interference process~$\mathbf{Y}(t)$, measured
at the origin of the two-dimensional plane in \cite[Fig. 1]{PinWin:J09}.
The spectral characteristics of $\mathbf{Y}(t)$ can be inferred from
the knowledge of its PSD.

\subsection{Power Spectral Density of the Aggregate Network Emission\label{sec:Power-Spectral-Density}}

\begin{figure}
\begin{centering}
\scalebox{1}{\psfrag{Xi(t)}{$\mathbf{X}_{i}(t)$}
\psfrag{Yi(t)}{$\mathbf{Y}_{i}(t)$}
\psfrag{h(t,tau)}{$\mathbf{h}_{i}(t,\tau)$}
\psfrag{e/R}{\large{$\frac{e^{\sigma G_{i}}}{R_{i}^{b}}$}}
\includegraphics{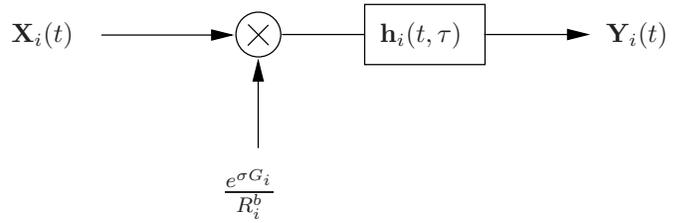}}
\par\end{centering}

\caption{\label{cap:PSDmdl}Channel model for spectral analysis.}

\end{figure}

The aggregate network emission at the probe receiver can be characterized
by the complex baseband random process~$\mathbf{Y}(t)$, defined
as\begin{equation}
\mathbf{Y}(t)=\sum_{i=1}^{\infty}\Yi,\label{eq:Yt-sumYi}\end{equation}
where $\Yi$ is the received process associated with each emitting
node~$i$. The signal~$\Yi$ can in turn be expressed for all time~$t$
as\begin{equation}
\Yi=\frac{e^{\sigma G_{i}}}{R_{i}^{b}}\int\mathbf{h}_{i}(t,\tau)\mathbf{X}_{i}(t-\tau)d\tau,\label{eq:Yit}\end{equation}
where $\mathbf{X}_{i}(t)$ is the complex baseband transmitted signal,
and $\mathbf{h}_{i}(t,\tau)$ is time-varying complex baseband impulse
response of the multipath channel associated with node~$i$. The
system model described by (\ref{eq:Yit}) is depicted in Fig.~\ref{cap:PSDmdl}.
It corresponds to a generalization of the model introduced in Part~I
of this paper, where we considered a two-dimensional modulation and
a flat Rayleigh fading channel. Since now we are interested in analyzing
the spectral properties of $\mathbf{Y}(t)$, we incorporate in the
model a generic transmitted waveform~$\mathbf{X}_{i}(t)$, not necessarily
associated with a two-dimensional modulation, as well as a generic
multipath channel~$\mathbf{h}_{i}(t,\tau)$, not necessarily associated
with flat Rayleigh fading. Also, since in this section we are only
interested in the aggregate emission of the network, we can ignore
the existence of the probe link depicted in \cite[Fig. 1]{PinWin:J09}.
In what follows we carry the analysis in complex baseband, although
it can be trivially translated to passband frequencies.

In the remainder of this paper, we consider that the transmitted signal~$\mathbf{X}_{i}(t)$
is a \emph{wide-sense stationary} (WSS) process, such that its autocorrelation
function has the form $R_{\mathbf{X}_{i}}(t_{1},t_{2})\triangleq\mathbb{E}\{\mathbf{X}_{i}^{*}(t_{1})\mathbf{X}_{i}(t_{2})\}=R_{\mathbf{X}}(\Delta t)$,
where $\Delta t=t_{2}-t_{1}$. We define the PSD of the process~$\mathbf{X}_{i}(t)$
as $\mathcal{S}_{\mathbf{X}}(f)\triangleq\mathcal{F}_{\Delta t\rightarrow f}\{R_{\mathbf{X}}(\Delta t)\}$.%
\footnote{We use $\mathcal{F}_{x\rightarrow y}\{\cdot\}$ to denote the Fourier
transform operator, where $x$ and $y$ represent the independent
variables in the original and transformed domains, respectively.%
} Since different nodes operate independently, the processes~$\mathbf{X}_{i}(t)$
are also independent for different~$i$, but the underlying second-order
statistics are the same (i.e.,~the autocorrelation function and the
PSD of $\mathbf{X}_{i}(t)$ do not depend on $i$). As we will show
in the case study of Section~\ref{sub:Psout-Plots}, if $\mathbf{X}_{i}(t)$
is a train of pulses with a uniformly distributed random delay (which
models the asynchronism between emitting nodes), then it is a WSS
process.

In terms of the multipath channel, we consider a \emph{wide-sense
stationary uncorrelated scattering} (WSSUS) channel~\cite{Bel:63,Par:00,Ste:92,Pro:00,Mol:B05},
so that the autocorrelation function of $\mathbf{h}_{i}(t,\tau)$
can be expressed as\begin{align*}
R_{\mathbf{h}_{i}}(t_{1},t_{2},\tau_{1},\tau_{2}) & \triangleq\mathbb{E}\{\mathbf{h}_{i}^{*}(t_{1},\tau_{1})\mathbf{h}_{i}(t_{2},\tau_{2})\}\\
 & =P_{\mathbf{h}}(\Delta t,\tau_{2})\delta(\tau_{2}-\tau_{1}),\end{align*}
for some function~$P_{\mathbf{h}}(\Delta t,\tau)$. Such channel
can be represented in the form of a densely-tapped delay line, as
a continuum of uncorrelated, randomly-scintillating scatterers having
WSS statistics. The functions~$\mathbf{h}_{i}(t,\tau)$ are considered
to be independent for different nodes~$i$, but the underlying second-order
statistics are the same (i.e.,~the autocorrelation function of $\mathbf{h}_{i}(t,\tau)$
does not depend on $i$). WSSUS channels are an important class of
practical channels which simultaneously exhibit wide-sense stationarity
in the time variable~$t$ and uncorrelated scattering in the delay
variable~$\tau$. They are the simplest non-degenerate channels which
exhibit both time and frequency fading, and also serve as a good model
for many radio channels.

We now wish to derive the PSD of the aggregate RF emission~$\mathbf{Y}(t)$
of the network, and with that purpose we introduce the following theorem.

\begin{theorem}[WSS and WSSUS Channels]\label{thm:WSSUS}Let $\mathbf{h}(t,\tau)$
denote the time-varying complex baseband impulse response of a multipath
channel, whose autocorrelation function is given by $R_{\mathbf{h}}(t_{1},t_{2},\tau_{1},\tau_{2})$.
Let $\mathbf{u}(t)$ denote the complex baseband WSS process which
is applied as input to the channel, and $\mathbf{z}(t)$ denote the
corresponding output process of the channel.
\begin{enumerate}
\item If the channel~$\mathbf{h}(t,\tau)$ is WSS, i.e., $R_{\mathbf{h}}(t_{1},t_{2},\tau_{1},\tau_{2})=R_{\mathbf{h}}(\Delta t,\tau_{1},\tau_{2})$,
then the output~$\mathbf{z}(t)$ is WSS and its PSD is given by%
\footnote{We use $\mathop*\limits ^{{\scriptscriptstyle x}}$ to denote the
convolution operation with respect to variable~$x$.  %
}\begin{equation}
\mathcal{S}_{\mathbf{z}}(f)=\int\int P_{\mathrm{s}}(\nu,\tau_{1},\tau_{2})|_{\nu=f}\mathop*\limits ^{{\scriptscriptstyle f}}\left[\mathcal{S}_{\mathbf{u}}(f)e^{j2\pi f(\tau_{1}-\tau_{2})}\right]d\tau_{1}d\tau_{2},\label{eq:Sz-WSS}\end{equation}
where $P_{\mathrm{s}}(\nu,\tau_{1},\tau_{2})\triangleq\mathcal{F}_{\Delta t\rightarrow\nu}\{R_{\mathbf{h}}(\Delta t,\tau_{1},\tau_{2})\}$,
and $\mathcal{S}_{\mathbf{u}}(f)$ is the PSD of $\mathbf{u}(t)$.
\item If the channel~$\mathbf{h}(t,\tau)$ is WSSUS, i.e., $R_{\mathbf{h}}(t_{1},t_{2},\tau_{1},\tau_{2})=P_{\mathbf{h}}(\Delta t,\tau_{2})\delta(\tau_{2}-\tau_{1})$
for some function~$P_{\mathbf{h}}(\Delta t,\tau)$, then the output~$\mathbf{z}(t)$
is WSS and its PSD is given by\begin{equation}
\mathcal{S}_{\mathbf{z}}(f)=\mathcal{D}_{\mathbf{h}}(\nu)|_{\nu=f}\mathop*\limits ^{{\scriptscriptstyle f}}\mathcal{S}_{\mathbf{u}}(f),\label{eq:Sz-WSSUS}\end{equation}
where $\mathcal{D}_{\mathbf{h}}(\nu)\triangleq\int P_{\mathrm{s}}(\nu,\tau)d\tau$
is the \emph{Doppler power spectrum} of the channel~$\mathbf{h}(t,\tau)$,
and $P_{\mathrm{s}}(\nu,\tau)\triangleq\mathcal{F}_{\Delta t\rightarrow\nu}\{P_{\mathbf{h}}(\Delta t,\tau)\}$
is the \emph{scattering function} of the channel~$\mathbf{h}(t,\tau)$.
\end{enumerate}
\end{theorem}

\begin{proof}See Appendix~\ref{sec:WSSUS-appendix} for a proof
and an intuitive interpretation of the theorem.\end{proof}

In the specific context of (\ref{eq:Yit}), the theorem implies that
$\Yi$ is WSS and thus the aggregate network emission~$\mathbf{Y}(t)$
is also WSS. Furthermore, the PSD of $\Yi$ is given by\begin{equation}
\SYi=\frac{e^{2\sigma G_{i}}}{R_{i}^{2b}}[\mathcal{D}_{\mathbf{h}}(f)*\mathcal{S}_{\mathbf{X}}(f)],\label{eq:SYi}\end{equation}
where $\mathcal{D}_{\mathbf{h}}(f)$ is the Doppler power spectrum
of the time-varying multipath channel~$\mathbf{h}_{i}(t,\tau)$,
and $\mathcal{S}_{\mathbf{X}}(f)$ is the PSD of the transmitted signal~$\mathbf{X}_{i}(t)$.
Because the processes~$\Yi$ associated with different emitting nodes~$i$
are statistically independent when conditioned on $\mathcal{P}$,
we can write\begin{equation}
\mathcal{S}_{\mathbf{Y}}(f)=\sum_{i=1}^{\infty}\SYi.\label{eq:SY-sumSYi}\end{equation}
Combining (\ref{eq:SYi}) and (\ref{eq:SY-sumSYi}), we obtain the
desired conditional PSD of the aggregate network emission~$\mathbf{Y}(t)$
as\begin{equation}
\mathcal{S}_{\mathbf{Y}}(f,\mathcal{P})=A\,[\mathcal{D}_{\mathbf{h}}(f)*\mathcal{S}_{\mathbf{X}}(f)],\label{eq:SY-final}\end{equation}
where $A$ was defined in (\ref{eq:A-sum}). Note that in (\ref{eq:SY-final})
we  explicitly indicated the conditioning of $\mathcal{S}_{\mathbf{Y}}$
on the random node positions and shadowing,~$\mathcal{P}$. Since
$\mathcal{S}_{\mathbf{Y}}(f,\mathcal{P})$ depends on $\mathcal{P}$,
it can be viewed, for a fixed~$f$, as a r.v.\ whose value is different
for each realization of $\mathcal{P}$.%
\footnote{$\mathcal{S}_{\mathbf{Y}}(f,\mathcal{P})$ is in fact a random process
whose sample paths evolve in \emph{frequency} instead of \emph{time}.
For each realization~$\mbox{$\mathcal{P}=\mathcal{P}_{0}$}$, we
obtain a sample path $\mathcal{S}_{\mathbf{Y}}(f,\mathcal{P}_{0})$
that is a function of $f$; for a fixed frequency~$f=f_{0}$, $\mathcal{S}_{\mathbf{Y}}(f_{0},\mathcal{P})$
is a r.v.%
}  Finally, we recall that $A$, when seen as a r.v., has the skewed
stable distribution given in (\ref{eq:stableA}).

\subsection{Spectral Outage Probability}

In the proposed quasi-static scenario, the PSD of the aggregate network
emission $\mathcal{S}_{\mathbf{Y}}(f,\mathcal{P})$ is a function
of the random node positions and shadowing,~$\mathcal{P}$. Then,
with some probability, $\mathcal{P}$ is such that the spectrum of
the aggregate emission is too high in some frequency band of interest,
thus causing an outage in that frequency band. This leads to the concept
of \emph{spectral outage probability} (SOP), which we denote by $P_{\textrm{out}}^{\textrm{s}}(f)$
and generally define as\begin{equation}
P_{\textrm{out}}^{\textrm{s}}(f)\triangleq\mathbb{P}_{\mathcal{P}}\{\mathcal{S}_{\mathbf{Y}}(f,\mathcal{P})>m(f)\},\label{eq:Psout-definition}\end{equation}
where $\mathcal{S}_{\mathbf{Y}}(f,\mathcal{P})$ is the random PSD
of the aggregate network emission~$\mathbf{Y}(t)$, and $m(f)$ is
some spectral mask determining the outage (or detection) threshold
at the receiver. The SOP is a frequency-dependent quantity and, in
the case of slow-varying positions~$\mathcal{P}$, is a more insightful
metric than the PSD averaged over $\mathcal{P}$. Note that this definition
is applicable in general to any emission model: the spectral outage
probability~$P_{\textrm{out}}^{\textrm{s}}(f)$ represents the probability
that the PSD of the aggregate network emission, measured at an arbitrary
location in the plane and at a particular frequency~$f$, exceeds
some predetermined mask. 

In commercial applications, the concept of SOP can provide a radically
different way to establish spectral regulations. Current regulations
and standards (e.g.,~FCC Part 15 or IEEE 802.11) impose a spectral
mask on the PSD \emph{at the transmitter}, and the type of mask often
depends on the environment in which the devices are operated (e.g.,~indoor
or outdoor). The purpose of this mask is to limit RF emissions generated
by a terminal, and to protect other services that operate in dedicated
bands (e.g.,~Global Positioning System, public safety, and cellular
systems). However, the \emph{transmitted} PSD is usually not representative
of the aggregate PSD \emph{at the victim receiver}, due to the random
propagation effects (multipath fading and shadowing) and the random
position of the emitting nodes. Thus, spectral regulations that are
based only on the transmitted PSD do not necessarily protect a victim
receiver against interference.

The approach proposed here is radically different, in the sense that
the spectral mask is defined at the victim receiver, not at the transmitter.
In effect, the mask~$m(f)$ introduced in (\ref{eq:Psout-definition})
represents the outage threshold with respect to the  \emph{accumulated
PSD at the receiver}, not the individual PSD at the transmitter (this
follows from the fact that $\mathcal{S}_{\mathbf{Y}}(f,\mathcal{P})$
is measured at an arbitrary location in the plane, where a probe receiver
could be located). Therefore, the received aggregate spectrum~$\mathcal{S}_{\mathbf{Y}}(f,\mathcal{P})$
and the corresponding~$P_{\textrm{out}}^{\textrm{s}}(f)$ can be
used to characterize and control the network's RF emissions more effectively,
since they not only consider the aggregate effect of all emitting
nodes at an arbitrary receiver location, but also incorporate the
random propagation effects and random node positions. Furthermore,
the use of different masks for indoor or outdoor environments is no
longer necessary, since the environment is already accounted for in
our model by parameters such as the amplitude loss exponent~$b$,
the spatial density~$\lambda$ of the emitting nodes, and the shadowing
coefficient~$\sigma$.

In military applications, on the other hand, the goal is to ensure
that the presence of the deployed network is not detected by the enemy.
If, for example, a surveillance network is to be deployed in enemy
territory, then the characterization of its aggregate emission is
essential for the design of a covert network with low probability
of detection. In such application, the function~$m(f)$ in (\ref{eq:Psout-definition})
can be interpreted as the frequency-dependent mask which determines
the \emph{detection threshold} (not the outage threshold as before).
In other words, if the aggregate spectral density~$\mathcal{S}_{\mathbf{Y}}(f,\mathcal{P})$
measured at a given location exceeds the mask~$m(f)$, then the presence
of the deployed network could be detected by the enemy.

For the signal model considered in this paper, $P_{\textrm{out}}^{\textrm{s}}(f)$
can be derived by substituting (\ref{eq:SY-final}) into the general
definition of SOP in (\ref{eq:Psout-definition}), leading to\begin{align}
P_{\textrm{out}}^{\textrm{s}}(f) & =\mathbb{P}\left\{ A>\frac{m(f)}{\mathcal{D}_{\mathbf{h}}(f)*\mathcal{S}_{\mathbf{X}}(f)}\right\} \nonumber \\
 & =1-F_{A}\left(\frac{m(f)}{\mathcal{D}_{\mathbf{h}}(f)*\mathcal{S}_{\mathbf{X}}(f)}\right),\label{eq:Psout-Fa}\end{align}
where $F_{A}(\cdot)$ is the cumulative distribution function (c.d.f.)
of the stable r.v.~$A$, whose distribution is given in (\ref{eq:stableA}).

\subsection{Numerical Results\label{sub:Psout-Plots}}

We now present a case study to quantify the spectral densities and
outage probabilities derived in the previous section. We also illustrate
their dependence on the various parameters involved, such as the transmitted
pulse shape, spectral mask, transmitted power, and spatial density
of the emitting nodes. For all numerical examples, we consider that
the emitting nodes employ a two-dimensional modulation  (e.g.,~$\mbox{$M$-PSK}$
or $\mbox{$M$-QAM}$), such that transmitted signal~$\mathbf{X}_{i}(t)$
can be written for all $t$ as\begin{equation}
\mathbf{X}_{i}(t)=\sum_{n=-\infty}^{+\infty}\mathbf{a}_{i,n}g(t-nT-D_{i}),\label{eq:Xit}\end{equation}
where the sequence~$\{\mathbf{a}_{i,n}\}_{n=-\infty}^{+\infty}$
represents the stream of complex symbols transmitted by node~$i$,
assumed to be independent identically distributed (i.i.d.) in $n$
and zero-mean, for simplicity; $g(t)$ is a real, baseband, unit-energy
shaping pulse, defined for all values of $t$; $T$ is the symbol
period; and $\mbox{$D_{i}\sim\mathcal{U}(0,T)$}$ is a random delay
representing the asynchronism between different emitting nodes. The
type of constellation employed by the emitting nodes is captured by
the statistics of the symbols~$\{\mathbf{a}_{i,n}\}$.%
\footnote{Note that each complex symbol~$\mathbf{a}_{i,n}=a_{i,n}e^{j\theta_{i,n}}$
can be represented in the IQ plane by a constellation point with amplitude~$a_{i,n}$
and phase~$\theta_{i,n}$.%
} Note that the process~$\mathbf{X}_{i}(t)$ in (\ref{eq:Xit}) is
WSS, as required by Theorem~\ref{thm:WSSUS}.%
\footnote{This can be shown in the following way: first, if we deterministically
set $D_{i}$ to zero in (\ref{eq:Xit}), the resulting process~$\widetilde{\mathbf{X}}_{i}(t)$
is wide-sense cyclostationary (WSCS) with period~$T$~\cite{PapPil:02};
then, since $\mbox{$\mathbf{X}_{i}(t)=\widetilde{\mathbf{X}}_{i}(t-D_{i})$}$,
with $\mbox{$D_{i}\sim\mathcal{U}(0,T)$}$ and independent of everything
else, it follows that $\mathbf{X}_{i}(t)$ is WSS.%
} The PSD of $\mathbf{X}_{i}(t)$ is then given by~\cite{SimHinLin:94,Win:J98,Win:J02}\begin{equation}
\mathcal{S}_{\mathbf{X}}(f)=P|G(f)|^{2},\label{eq:Sx-example}\end{equation}
where $P=\mathbb{E}\{|\mathbf{a}_{i,n}|^{2}\}/T$ is the power transmitted
by each emitting node, and $G(f)=\mathcal{F}\{g(t)\}$.

In terms of the multipath channel, we consider for simplicity that
$\mathbf{h}(t,\tau)$ is time-invariant such that it does not introduce
any Doppler shifts, i.e.,~$\mathcal{D}_{\mathbf{h}}(\nu)=\delta(\nu)$.%
\footnote{For typical node speeds or channel fluctuations, the frequencies of
the Doppler shifts are on the order of few KHz. As a consequence,
when the considered $\mathbf{X}_{i}(t)$ is an ultrawideband signal,
$\mathcal{D}_{\mathbf{h}}(\nu)$ can be well approximated by a Dirac-delta
function.%
} Substituting the expressions for $\mathcal{S}_{\mathbf{X}}(f)$ and
$\mathcal{D}_{\mathbf{h}}(\nu)$ in (\ref{eq:Psout-Fa}), we obtain
the SOP as\begin{equation}
P_{\textrm{out}}^{\textrm{s}}(f)=1-F_{A}\left(\frac{m(f)}{P|G(f)|^{2}}\right).\label{eq:Psout-example}\end{equation}

\begin{figure}
\begin{centering}
\subfigure[PSD of the individual transmitted signal versus frequency (bottom curves), for various pulse shapes~$g(t)$. The square and Hanning pulses are normalized so that the transmitted signals have the same power~$P$. The piecewise-constant spectral mask~$m(f)$ (top curve) determines the outage threshold at the receiver.]{
  \label{cap:psd-pulse-plot}
  \scalebox{0.45}{
    \psfrag{frequency f (MHz)}{\Large{\sf{frequency $f$ (MHz)}}}
    \psfrag{power spectral density (dBm/Hz)}{\Large{\sf{power spectral density (dBm/Hz)}}}
    \includegraphics{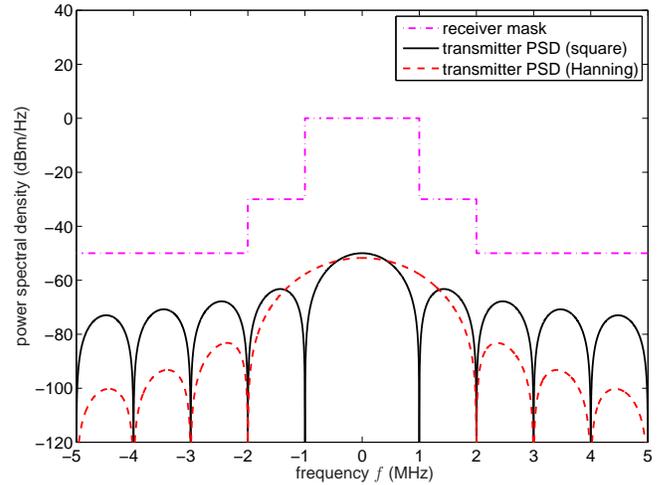}
  }
}
\par\end{centering}

\begin{centering}
\subfigure[Spectral outage probability~$P_{\textrm{out}}^{\textrm{s}}(f)$ versus frequency, for the piecewise-constant mask~$m(f)$ shown in (a).]{
   \label{cap:psout-pulse-plot}
   \scalebox{0.45}{
     \psfrag{frequency f (MHz)}{\Large{\sf{frequency $f$ (MHz)}}}
     \psfrag{Psout}{\LARGE{$P_{\textrm{out}}^{\textrm{s}}(f)$}}
     \includegraphics{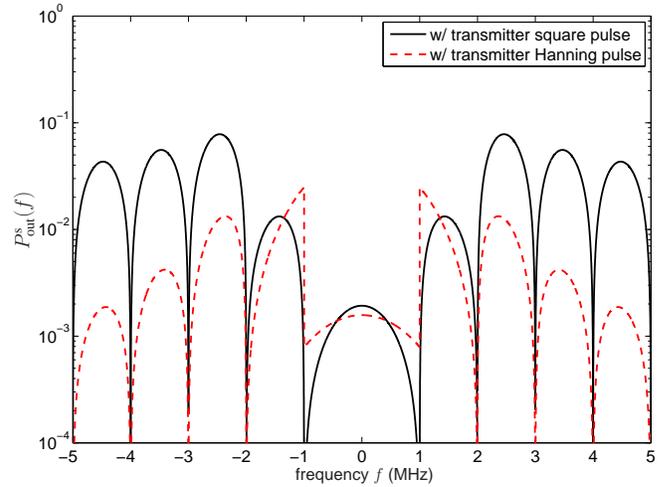}
  }
}
\par\end{centering}

\caption{\label{cap:effect-of-pulse}Effect of the transmitted baseband pulse
shape~$g(t)$ on the PSD and the outage probability~$P_{\textrm{out}}^{\textrm{s}}(f)$
($\mbox{$P=10\,\textrm{dBm}$}$, $\mbox{$T=10^{-6}\,\textrm{s}$}$,
$\mbox{$\lambda=0.1\,\textrm{m}^{-2}$}$, $\mbox{$b=2$}$, $\sigma_{\textrm{dB}}=10$).}

\end{figure}

Figure~\ref{cap:effect-of-pulse} shows that for a fixed spectral
mask~$m(f)$, the SOP can be highly dependent on the pulse shape~$g(t)$
(e.g.,~square or Hanning pulse). In fact, $P_{\textrm{out}}^{\textrm{s}}(f)$
is a nonlinear function of $|G(f)|$, where the nonlinearity is determined
in part by the c.d.f.~$F_{A}(\cdot)$ of the stable r.v.~$A$, as
shown in (\ref{eq:Psout-example}). Thus, the SOP can be used as a
criterion for designing the pulse shape: for example, we may wish
to determine the baseband pulse~$g(t)$ and transmitted power~$P$
such that $\mbox{$\max_{f}P_{\textrm{out}}^{\textrm{s}}(f)\leq p^{*}$}$,
where $p^{*}$ is some target outage probability which must be satisfied
at all frequencies.

\begin{figure}
\begin{centering}
\subfigure[Plot of various spectral masks~$m(f)$ which define the outage threshold at the receiver (top curves). Also shown is the PSD of the individual transmitted signal versus frequency (bottom curve).]{
  \label{cap:psd-mask-plot}
  \scalebox{0.45}{
    \psfrag{frequency f (MHz)}{\Large{\sf{frequency $f$ (MHz)}}}
    \psfrag{power spectral density (dBm/Hz)}{\Large{\sf{power spectral density (dBm/Hz)}}}
    \includegraphics{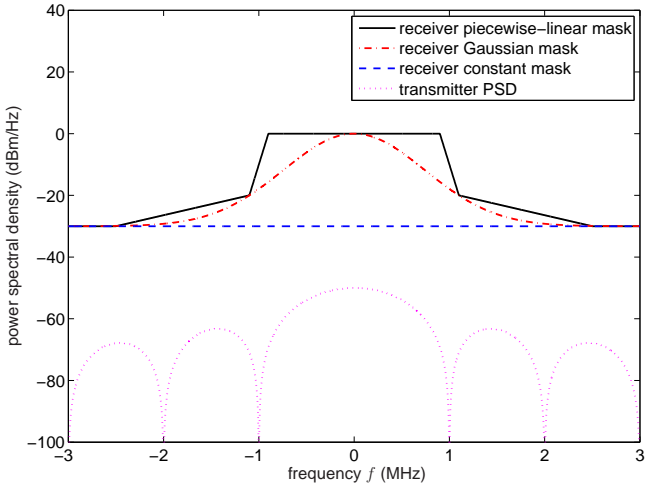}
  }
}
\par\end{centering}

\begin{centering}
\subfigure[Spectral outage probability~$P_{\textrm{out}}^{\textrm{s}}(f)$ versus frequency, for the various masks~$m(f)$ shown in (a).]{
   \label{cap:psout-mask-plot}
   \scalebox{0.45}{
     \psfrag{frequency f (MHz)}{\Large{\sf{frequency $f$ (MHz)}}}
     \psfrag{Psout}{\LARGE{$P_{\textrm{out}}^{\textrm{s}}(f)$}}
     \includegraphics{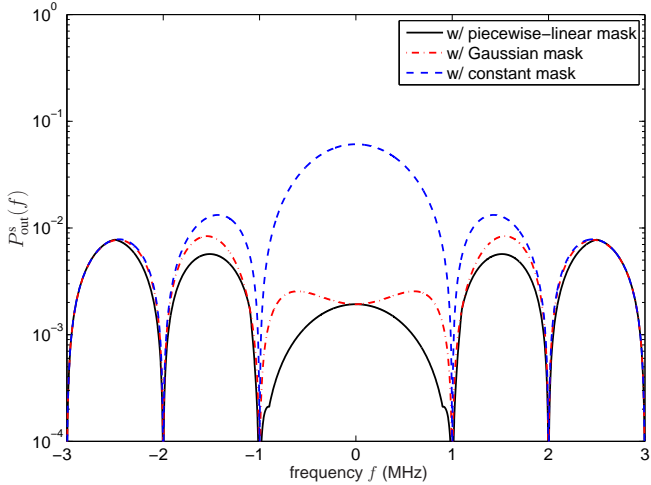}
  }
}
\par\end{centering}

\caption{\label{cap:effect-of-mask}Effect of the spectral mask shape~$m(f)$
on the outage probability~$P_{\textrm{out}}^{\textrm{s}}(f)$ (square
$g(t)$, $\mbox{$P=10\,\textrm{dBm}$}$, $\mbox{$T=10^{-6}\,\textrm{s}$}$,
$\mbox{$\lambda=0.1\,\textrm{m}^{-2}$}$, $\mbox{$b=2$}$, $\sigma_{\textrm{dB}}=10$).}

\end{figure}

Figure~\ref{cap:effect-of-mask} shows that for a fixed pulse shape~$g(t)$,
$P_{\textrm{out}}^{\textrm{s}}(f)$ can significantly depend on the
spectral mask~$m(f)$ (e.g.,~piecewise-linear, Gaussian, or constant
mask). Since $P_{\textrm{out}}^{\textrm{s}}(f)$ accounts for both
$G(f)$ and $m(f)$, it quantifies the compatibility of the transmitted
pulse shape with the spectral restrictions imposed through $m(f)$.

\begin{figure}
\begin{centering}
\scalebox{0.45}{\psfrag{Psout}{\LARGE{$P_{\textrm{out}}^{\textrm{s}}(f)$}}
\psfrag{frequency f (MHz)}{\Large{\sf{frequency $f$ (MHz)}}}
\psfrag{P=20 dBm}{\Large{$P=\mathsf{20\;dBm}$}}
\psfrag{P=10 dBm}{\Large{$P=\mathsf{10\;dBm}$}}
\psfrag{P=0 dBm}{\Large{$P=\mathsf{0\;dBm}$}}\includegraphics{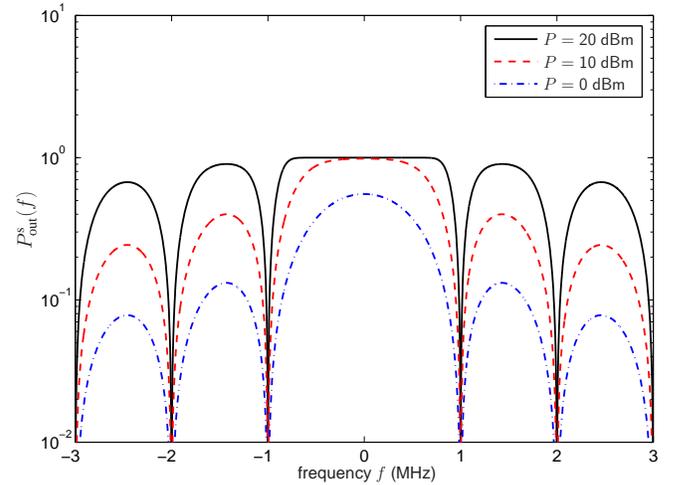}}
\par\end{centering}

\caption{\label{cap:psout-power-plot}Spectral outage probability~$P_{\textrm{out}}^{\textrm{s}}(f)$
versus frequency, for various transmitted powers~$P$ (square $g(t)$,
$\mbox{$T=10^{-6}\,\textrm{s}$}$, $\mbox{$\lambda=0.1\,\textrm{m}^{-2}$}$,
$\mbox{$b=2$}$, $\sigma_{\textrm{dB}}=10$, $\mbox{$m(f)=-60\,\textrm{dBm/Hz}$}$).}

\end{figure}

\begin{figure}
\begin{centering}
\scalebox{0.45}{\psfrag{Psout(f=0)}{\LARGE{$P_{\textrm{out}}^{\textrm{s}}(f=0)$}}
\psfrag{transmitted power P (dBm)}{\Large{\sf{transmitted power $P$ (dBm)}}}
\psfrag{la=0.01,0.1,1}{\LARGE{$\mathsf{\lambda=0.01,0.1,1\, m^{-2}}$}}\includegraphics{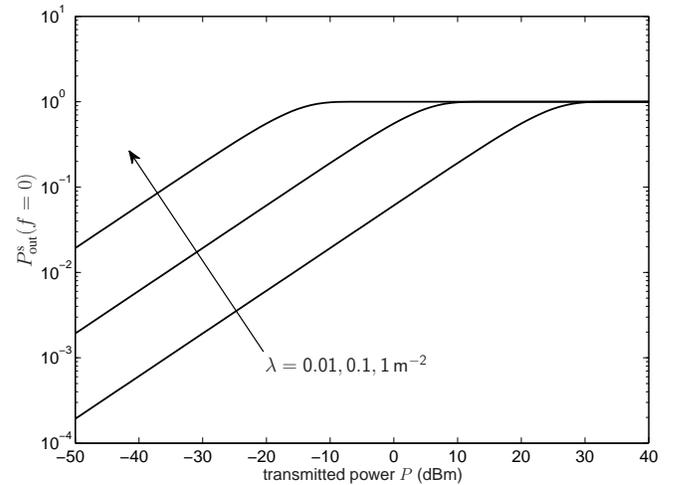}}
\par\end{centering}

\caption{\label{cap:psout-lambda-plot}Spectral outage probability~$P_{\textrm{out}}^{\textrm{s}}(f)$,
evaluated at $f=0$, versus transmitted power~$P$, for various spatial
densities~$\lambda$ of the emitting nodes (square $g(t)$, $\mbox{$T=10^{-6}\,\textrm{s}$}$,
$\mbox{$b=2$}$, $\sigma_{\textrm{dB}}=10$, $\mbox{$m(f)=-60\,\textrm{dBm/Hz}$}$).}

\end{figure}

Figures~\ref{cap:psout-power-plot} and \ref{cap:psout-lambda-plot}
illustrate, respectively, the dependence of the outage probability~$P_{\textrm{out}}^{\textrm{s}}(f)$
on the transmitted power~$P$ and spatial density~$\lambda$ of
the emitting nodes. Specifically, as $P$ or $\lambda$ increase,
the aggregate network emission becomes stronger, and thus $P_{\textrm{out}}^{\textrm{s}}(f)$
deteriorates at all frequencies, approaching the maximum value of
1.

\subsection{Generalizations\label{sec:Generalizations}}

We now extend the results to an heterogeneous scenario with $K$ different
networks, where a given emitting node belongs to the network~$k\in\{1\ldots K\}$
with probability~$p_{k}$, independently of everything else. Using
the splitting property of Poisson processes~\cite{BerTsi:02}, we
know the emitting nodes from each network~$k$ form a spatial Poisson
process, which is independent of the processes of other networks and
has spatial density~$\lambda_{k}=\lambda p_{k}$. Therefore, we can
write the aggregate emission from all nodes in all networks as\begin{equation}
\mathbf{Y}(t)=\sum_{k=1}^{K}\mathbf{Y}_{k}(t),\label{eq:Yt-sumYki}\end{equation}
where $\mathbf{Y}_{k}(t)=\sum_{i=1}^{\infty}\mathbf{Y}_{k,i}(t)$
is the aggregate emission from the individual network~$k$, and\begin{equation}
\mathbf{Y}_{k,i}(t)=\frac{e^{\sigma G_{k,i}}}{R_{k,i}^{b}}\int\mathbf{h}_{k,i}(t,\tau)\mathbf{X}_{k,i}(t-\tau)d\tau,\quad k=1\ldots K,\label{eq:Yki(t)-sum}\end{equation}
where $\mathbf{X}_{k,i}(t)$ and $\mathbf{h}_{k,i}(t,\tau)$ are,
respectively, the transmitted signal and the impulse response of the
multipath channel associated with node~$i$ from network~$k$. We
consider that $\mathbf{X}_{k,i}(t)$ and $\mathbf{h}_{k,i}(t,\tau)$
are independent in both $k$ and $i$. Then, the aggregate emission~$\mathbf{Y}_{k}(t)$
is also independent for different networks~$k$ when conditioned
on the positions~$\mathcal{P}$, and thus $\mathcal{S}_{\mathbf{Y}}(f)=\sum_{k=1}^{K}\mathcal{S}_{\mathbf{Y}_{k}}(f).$
We can generalize (\ref{eq:SY-final}) and write the conditional PSD
of the aggregate emission~$\mathbf{Y}(t)$ in this heterogeneous
scenario as \begin{equation}
\mathcal{S}_{\mathbf{Y}}(f,\mathcal{P})=\sum_{k=1}^{K}A_{k}\,[\mathcal{D}_{\mathbf{h}_{k}}(f)*\mathcal{S}_{\mathbf{X}_{k}}(f)],\label{eq:SY-gen}\end{equation}
where $\mathcal{D}_{\mathbf{h}_{k}}(f)$ and $\mathcal{S}_{\mathbf{X}_{k}}(f)$
are, respectively, the Doppler power spectrum and the PSD of the transmitted
signal associated with network~$k$; and the r.v.'s~$\{A_{k}\}$
are i.i.d.~in $k$ and given by\[
A_{k}=\sum_{i=1}^{\infty}\frac{e^{2\sigma G_{k,i}}}{R_{k,i}^{2b}}.\]
In (\ref{eq:SY-gen}), we have explicitly indicated the conditioning
of $\mathcal{S}_{\mathbf{Y}}$ on the random node positions and shadowing~$\mathcal{P}$.
Since $A_{k}$ depends on $\mathcal{P}$, it can be seen as a r.v.\ whose
distribution is skewed stable,\begin{equation}
A_{k}\sim\mathcal{S}\left(\alpha_{A}=\frac{1}{b},\:\beta_{A}=1,\:\gamma_{A}=\lambda_{k}\pi C_{1/b}^{-1}e^{2\sigma^{2}/b^{2}}\right).\label{eq:stableA-gen}\end{equation}

\section{Summary\label{sec:Summary}}

This two-part paper investigates a mathematical model for communication
subject to both network interference and AWGN, where the spatial distribution
of the nodes is captured by a Poisson field in the two-dimensional
plane. We specifically address the cases of slow and fast-varying
node positions, as well as homogeneous and heterogeneous networks,
in a realistic wireless environment subject to path loss, multipath
fading and shadowing. In Part~I, we determined the statistical distribution
of the aggregate interference at the output of a conventional linear
receiver, which leads directly to the characterization of the error
performance (in terms of outage and average probabilities). 

In this second part, we characterize the capacity of the link when
subject to both network interference and noise, and derive the PSD
of the aggregate RF emission of the network. Then, we put forth the
concept of spectral outage probability (SOP), and describe some possible
applications, namely the establishment of spectral regulations and
the design of covert military networks. In particular, the SOP can
be used as a criterion for designing pulse shapes or controlling interference
in wireless networks, and as a measure of the network's covertness.
Our framework clearly shows how the aggregate network emission can
be characterized in terms of important network parameters, thereby
providing insights that may be of value to the network designer.

\appendices

\section{Derivation of Theorem~\ref{thm:WSSUS}\label{sec:WSSUS-appendix}}

The derivation of Theorem~\ref{thm:WSSUS} relies on the general
theory of linear time-varying systems and Bello system functions~\cite{Bel:63,Par:00,Ste:92,Pro:00}.
Let $\mathbf{h}(t,\tau)$ denote a time-varying complex baseband impulse
response of a multipath channel. When the complex baseband process~$\mathbf{u}(t)$
is applied as input to the channel, the output process~$\mathbf{z}(t)$
is given by the integral\[
\mathbf{z}(t)=\int\mathbf{h}(t,\tau)\mathbf{u}(t-\tau)d\tau.\]
We define the autocorrelation function of the input~$\mathbf{u}(t)$
as $R_{\mathbf{u}}(t_{1},t_{2})\triangleq\mathbb{E}\{\mathbf{u}^{*}(t_{1})\mathbf{u}(t_{2})\}$,
and the autocorrelation function of the channel~$\mathbf{h}(t,\tau)$
as $R_{\mathbf{h}}(t_{1},t_{2},\tau_{1},\tau_{2})\triangleq\mathbb{E}\{\mathbf{h}^{*}(t_{1},\tau_{1})\mathbf{h}(t_{2},\tau_{2})\}$.
The autocorrelation of the output~$\mathbf{z}(t)$ is generally given
by\begin{equation}
R_{\mathbf{z}}(t_{1},t_{2})=\int\int R_{\mathbf{h}}(t_{1,}t_{2},\tau_{1},\tau_{2})R_{\mathbf{u}}(t_{1}-\tau_{1},t_{2}-\tau_{2})d\tau_{1}d\tau_{2}.\label{eq:Rz-general}\end{equation}
Since the input process~$\mathbf{u}(t)$ is WSS, $R_{\mathbf{u}}(t_{1,}t_{2})=R_{\mathbf{u}}(\Delta t)$,
where $\Delta t=t_{2}-t_{1}$.

We first consider a WSS channel~$\mathbf{h}(t,\tau)$ such that $R_{\mathbf{h}}(t_{1},t_{2},\tau_{1},\tau_{2})=R_{\mathbf{h}}(\Delta t,\tau_{1},\tau_{2})$.
Then, we can rewrite $R_{\mathbf{z}}(t_{1},t_{2})$ in (\ref{eq:Rz-general})
as\begin{align}
R_{\mathbf{z}}(t_{1},t_{2}) & =\int\int R_{\mathbf{h}}(\Delta t,\tau_{1},\tau_{2})R_{\mathbf{u}}(\Delta t+\tau_{1}-\tau_{2})d\tau_{1}d\tau_{2}\nonumber \\
 & \triangleq R_{\mathbf{z}}(\Delta t).\label{eq:Rz-WSS}\end{align}
Since $R_{\mathbf{z}}(t_{1},t_{2})$ is a function only of $\Delta t$,
the output~$\mathbf{z}(t)$ is also WSS. The PSD of $\mathbf{z}(t)$
can be written as\begin{align}
\mathcal{S}_{\mathbf{z}}(f) & =\mathcal{F}_{\Delta t\rightarrow f}\{R_{\mathbf{z}}(\Delta t)\}\nonumber \\
 & =\int\int\left[\int R_{\mathbf{h}}(\Delta t,\tau_{1},\tau_{2})R_{\mathbf{u}}(\Delta t+\tau_{1}-\tau_{2})\right.\nonumber \\
 & \times\left.e^{-j2\pi f(\Delta t)}d(\Delta t)\right]d\tau_{1}d\tau_{2}\nonumber \\
 & =\int\int P_{\mathrm{s}}(\nu,\tau_{1},\tau_{2})|_{\nu=f}\mathop*\limits ^{{\scriptscriptstyle f}}\left[\mathcal{S}_{\mathbf{u}}(f)e^{j2\pi f(\tau_{1}-\tau_{2})}\right]d\tau_{1}d\tau_{2},\label{eq:Sz-WSS-repeated}\end{align}
where $P_{\mathrm{s}}(\nu,\tau_{1},\tau_{2})\triangleq\mathcal{F}_{\Delta t\rightarrow\nu}\{R_{\mathbf{h}}(\Delta t,\tau_{1},\tau_{2})\}$,
and $\mathcal{S}_{\mathbf{u}}(f)$ is the PSD of $\mathbf{u}(t)$.
This is the result in Theorem~\ref{thm:WSSUS}, eq.~(\ref{eq:Sz-WSS}).

We now further constrain the channel~$\mathbf{h}(t,\tau)$ to be
WSSUS such that $R_{\mathbf{h}}(t_{1},t_{2},\tau_{1},\tau_{2})=P_{\mathbf{h}}(\Delta t,\tau_{2})\delta(\tau_{2}-\tau_{1}),$
for some function~$P_{\mathbf{h}}(\Delta t,\tau)$. Then, $R_{\mathbf{z}}(t_{1},t_{2})$
in (\ref{eq:Rz-WSS}) can be further simplified as follows:\begin{align}
R_{\mathbf{z}}(t_{1},t_{2}) & =\int\int P_{\mathbf{h}}(\Delta t,\tau_{2})\delta(\tau_{2}-\tau_{1})R_{\mathbf{u}}(\Delta t+\tau_{1}-\tau_{2})d\tau_{1}d\tau_{2}\nonumber \\
 & =\int P_{\mathbf{h}}(\Delta t,\tau)R_{\mathbf{u}}(\Delta t)d\tau\nonumber \\
 & =R_{\mathbf{u}}(\Delta t)\int P_{\mathbf{h}}(\Delta t,\tau)d\tau\triangleq R_{\mathbf{z}}(\Delta t).\label{eq:Rz-WSSUS}\end{align}
The output~$\mathbf{z}(t)$ is therefore WSS, and its PSD can be
written as\begin{align}
\mathcal{S}_{\mathbf{z}}(f) & =\mathcal{F}_{\Delta t\rightarrow f}\{R_{\mathbf{z}}(\Delta t)\}\nonumber \\
 & =\mathcal{F}_{\Delta t\rightarrow f}\{R_{\mathbf{u}}(\Delta t)\}*\mathcal{F}_{\Delta t\rightarrow f}\left\{ \int P_{\mathbf{h}}(\Delta t,\tau)d\tau\right\} \nonumber \\
 & =\mathcal{S}_{\mathbf{u}}(f)\mathop*\limits ^{{\scriptscriptstyle f}}\int P_{\mathrm{s}}(\nu,\tau)|_{\nu=f}d\tau,\label{eq:Sz-WSSUS-repeated}\end{align}
where $P_{\mathrm{s}}(\nu,\tau)\triangleq\mathcal{F}_{\Delta t\rightarrow\nu}\{P_{\mathbf{h}}(\Delta t,\tau)\}$
is known as the \emph{scattering function} of the channel~$\mathbf{h}(t,\tau)$.
It provides a measure of the average power output of the channel as
a function of the delay~$\tau$ and the Doppler shift~$\nu$. Furthermore,
if we define the \emph{Doppler power spectrum} of the channel as $\mathcal{D}_{\mathbf{h}}(\nu)\triangleq\int P_{\mathrm{s}}(\nu,\tau)d\tau$,
then (\ref{eq:Sz-WSSUS-repeated}) can be succinctly written as\begin{equation}
\mathcal{S}_{\mathbf{z}}(f)=\mathcal{S}_{\mathbf{u}}(f)\mathop*\limits ^{{\scriptscriptstyle f}}\mathcal{D}_{\mathbf{h}}(\nu)|_{\nu=f},\label{eq:Sz-simple}\end{equation}
which is the result in Theorem~\ref{thm:WSSUS}, eq.~(\ref{eq:Sz-WSSUS}).

From (\ref{eq:Sz-simple}), we conclude that $\mathcal{S}_{\mathbf{z}}(f)$
depends on the \emph{Doppler power spectrum} of the channel, $\int P_{\mathrm{s}}(\nu,\tau)d\tau$,
but not on its \emph{power delay profile}~$\int P_{\mathrm{s}}(\nu,\tau)d\nu$.
This is intuitively satisfying since all delayed replicas of the WSS
process~$\mathbf{u}(t)$ have the same PSD. Furthermore, if the channel~$\mathbf{h}(t,\tau)$
is time-invariant, then $\mathcal{D}_{\mathbf{h}}(\nu)=\delta(\nu)$
and thus $\mathcal{S}_{\mathbf{z}}(f)=\mathcal{S}_{\mathbf{u}}(f)$,
i.e.,~the channel does not affect the PSD of the input. On the other
hand, if the channel is time-varying in such a way that it introduces
a Doppler shift of $f_{0}$~Hz, then $\mathcal{D}_{\mathbf{h}}(\nu)=\delta(\nu-f_{0})$
and thus $\mathcal{S}_{\mathbf{z}}(f)=\mathcal{S}_{\mathbf{u}}(f-f_{0})$,
i.e.,~the output PSD is simply the input PSD shifted by $f_{0}$~Hz,
as expected.

\section*{Acknowledgements}

\noindent The authors would like to thank L.~A.~Shepp, R.~A.~Scholtz,
L.~Greenstein, J.~H.~Winters, G.~J.~Foschini, M.~Chiani, and
A.~Giorgetti for their helpful suggestions.

\bibliographystyle{../bibtex/IEEEtran}
\bibliography{../bibtex/IEEEabrv,../bibtex/StringDefinitions,../bibtex/WGroup,../bibtex/BiblioCV}

\end{document}